\documentclass{article}
\pdfoutput=1

\usepackage{arxiv}

\usepackage[utf8]{inputenc} % allow utf-8 input
\usepackage[T1]{fontenc}    % use 8-bit T1 fonts
\usepackage{amsfonts}       % blackboard math symbols
\usepackage{nicefrac}       % compact symbols for 1/2, etc.
\usepackage{microtype}      % microtypography
\usepackage{graphicx}
\usepackage{amsmath}
\usepackage{subfigure}

\title{Wi-Fi Passive Person Re-Identification based on Channel State Information}

\author{
  Danilo Avola\\
  Department of Computer Science\\
  Sapienza University of Rome\\
  Via Salaria 113, Roma, RM 00198 \\
  \texttt{avola@di.uniroma1.it} \\
    \And
    Marco Cascio\thanks{Corresponding Author \newline E-mail address: cascio@di.uniroma1.it (Marco Cascio)} \\
    Department of Computer Science\\
    Sapienza University of Rome\\
    Via Salaria 113, Roma, RM 00198 \\
    \texttt{cascio@di.uniroma1.it} \\
    \AND
    Luigi Cinque\\
    Department of Computer Science\\
    Sapienza University of Rome\\
    Via Salaria 113, Roma, RM 00198 \\
    \texttt{cinque@di.uniroma1.it} \\
    \And
    Daniele Pannone\\
    Department of Computer Science\\
    Sapienza University of Rome\\
    Via Salaria 113, Roma, RM 00198 \\
    \texttt{pannone@di.uniroma1.it} \\
}

\begin{document}
\maketitle

\begin{abstract}
With the increasing need for wireless data transfer, Wi-Fi networks have rapidly grown in recent years providing high throughput and easy deployment. Nowadays, Access Points (APs) can be found easily wherever we go, therefore Wi-Fi sensing applications have caught a great deal of interest from the research community. Since human presence and movement influence the Wi-Fi signals transmitted by APs, it is possible to exploit those signals for person  Re-Identification (Re-ID) task. Traditional techniques for Wi-Fi sensing applications are usually based on the Received Signal Strength Indicator (RSSI) measurement. However, recently, due to the RSSI instability, the researchers in this field propose Channel State Information (CSI) measurement based methods. In this paper we explain how changes in Signal Noise Ratio (SNR), obtained from CSI measurements, combined with Neural Networks can be used for person Re-ID achieving remarkable preliminary results. Due to the lack of available public data in the current state-of-the-art to test such type of task, we acquired a dataset that properly fits the aforementioned task.
\end{abstract}

% keywords can be removed

\section{Introduction}\label{01}
With recent developments in the Internet of Things (IoT) technology, Wi-Fi infrastructures have become common in both public and private areas. Radio signals transmitted by several Access Points (APs), more than just connecting to the Internet, can be exploited for various Wi-Fi sensing tasks \cite{7875144, 8057278, 8399492}, including person Identification (ID) \cite{Korany:2019:XUW:3300061.3345437} and Re-Identification (Re-ID). Person Re-ID is the task aimed to recognize an already known person identity by comparing a probe (i.e., person of interest) to a gallery of candidates and, generally, it is performed by exploiting visual data requiring specific sensors in specific locations. Considering a real-world indoor surveillance scenario, for example, several cameras strategically located at different rooms need to be used. Moreover, visual-based methods have to face the well-known vision challenges related to occlusions, illumination changes and similarity/appearance issues among human subjects. Differently, Wi-Fi based techniques could be new non-invasive and alternative ways to perform the Re-ID task, even more robust to aforementioned vision related challenges. Notice that Wi-Fi signals pass-through obstructions, for example. In addition, whereas the use of cameras may violate human rights or subject's privacy, radio signals do not suffer from such issues. 

Either human presence or movement between Wi-Fi transmitters and receivers influence the radio signals transmitted \cite{8514811}, thus Wi-Fi based sensing applications are feasible by analysing both the Channel State Information (CSI) measurement \cite{Halperin:2011:TRG:1925861.1925870} from the Wi-Fi signal and the features that are possible to extract by processing it (e.g., signal amplitude or phase). Specifically, the CSI describes the characteristics of the wireless communication channel and, in detail, such characteristics represent how the wireless signal propagate from the transmitter to the receiver at specific carrier frequencies along multiple paths. This measure allows inferring many detail about the signal, including scattering or attenuation based on path length. Methods based on wireless signals can be categorised in Active (i.e., Device-Based) \cite{MOHAMMED201650, 8314747} and Passive (i.e., Device-Free) \cite{7458186, 8354833} approaches. Whereas the former perform Wi-Fi sensing tasks by exploiting a device worn by the human subject, the latter do not. Such techniques, especially the Device-Free ones, add novelty to the state-of-the-art and they found application in a wide range of fields such as surveillance, security, healthcare, monitoring, imaging, human-computer interaction and many others.

In this paper, we propose a Device-Free Wi-Fi method able to re-identify person identities in different indoor environments. Our approach, by using a neural network for re-identification, exploits CSI measurements to estimate the Signal Noise Ratio (SNR) of received wireless signals. Through preliminary experiments, has been verified that human biometrics are embedded even in SNR estimated from such signals. In addition, due to lack of publicly available datasets, we also collected data to validate the proposed method. As far as we know, in current state-of-the-art we are the first in exploiting the CSI for the Re-ID task. Moreover, our work demonstrates that the use of commercial Wi-Fi devices is possible to capture human biometrics for people Re-ID in real-world indoor scenarios.

The paper is structured as follows. In Section \ref{02}, a brief overview about the current literature concerning Wi-Fi human-related sensing applications is reported. In Section \ref{03}, the proposed method architecture is described in detail, including CSI measurement and SNR estimation. In Section \ref{04}, starting with the description of the datasets collected to fit the Wi-Fi person Re-ID task, the obtained experimental results are discussed. Finally, in Section \ref{05} the paper is concluded.

\section{State-of-the-art}\label{02}
Tipically, for Wi-Fi sensing systems development the most used measurement is the Received Signal Strength Indicator (RSSI), which expresses the relative signal quality by indicating the power level being received after any possible loss at the antenna and cable levels.
The authors of \cite{5534913} reported that human presence influences the performance of wireless communications, causing significantly RSSI fluctuation in both line-of-sight (LOS) and non-line-of-sight (NLOS) conditions. Indeed, such measurement is widely used for human localization \cite{7529124,8263603,8412766}, detection \cite{7733160,8263558} and pose estimation \cite{7218525} tasks. In \cite{8412766}, the RSSI has been even used for person identification by exploiting a proximity based algorithm in a Device-Based setting. Wireless sensors have been placed on specific environmental objects, the user wearing the device that features the highest RSSI is identified as the one actually interacting with the surrounding area. However, even if the RSSI is easy to obtain, it is an unstable and noisy measure unable to capture the real changes in the signal, due to multi-path and fading effects, offering limited performance \cite{Xiao:2016:SWI:2966278.2933232}. 

More and more researchers recently start to use the Channel State Information for its great signal processing capability. Notice that, the CSI allows the use of both signal processing and computer vision algorithms. In \cite{7458186}, the authors propose an automatic Wi-Fi system for human detection and pose estimation by exploiting CSI measurements to extract the Wi-Fi signal amplitude. In \cite{7807197}, instead, the authors introduce the theoretical analysis of the sensing capability of Wi-Fi signals, by defining a Fresnel zone \cite{0072561912} model for human breathing recognition and pose estimation. Differently, the authors of \cite{8004463}, perform Wi-Fi human localization and detection tasks by using radio images features. Specifically, the CSI measurements from different communication channels are used to obtain radio images from which visual features are extracted. The CSI has also been used for person identification, and existing works are mostly based on biometric signatures. The authors in \cite{Chen:2017:RMD:3139486.3130906} propose a passive method based on an accurate gait analysis obtained by fusing CSI and footstep sound measurements. Also in \cite{Wang:2016:GRU:2971648.2971670}, gait patterns are captured to recognize people. The authors characterize walking patterns by using spectrograms from CSI. Even in  \cite{Shi:2017:SUA:3084041.3084061}, the user-authentication is obtained by leveraging on discriminative features extracted from CSI measurements of prevalent WiFi signals captured during daily activities, including walking. Again, in \cite{7803604}, the authors exploit structural biometric features. They perform statistical analysis of the received Channel Frequency Response (i.e., the frequency-domain CSI) amplitude and phase.
Differently, in \cite{7955011}, to achieve user-authentication in secure environments, the presence of a spoofer is detected passively by examining the temporal correlation of CSI measurements. In this case, the authors remark the expected better results compared with existing approaches based on RSSI.

In current literature, Wi-Fi based person Re-ID works exploiting CSI measurements do not exist. Existing methods are mostly based on visual information extracted by processing images or video sequences \cite{ZHOU2018739, 8100017, 10.1007/978-3-030-29891-3_41, 10.1007/978-3-030-01216-8_18}. The publicly available datasets used by those approaches do not provide wireless signals data, therefore we cannot provide direct comparisons with state-of-the-art. We collected data by ourselves to build an evaluation benchmark for the proposed method and we reported the results.

\begin{figure}[t!]
	\centering
	\includegraphics[width=0.5\columnwidth]{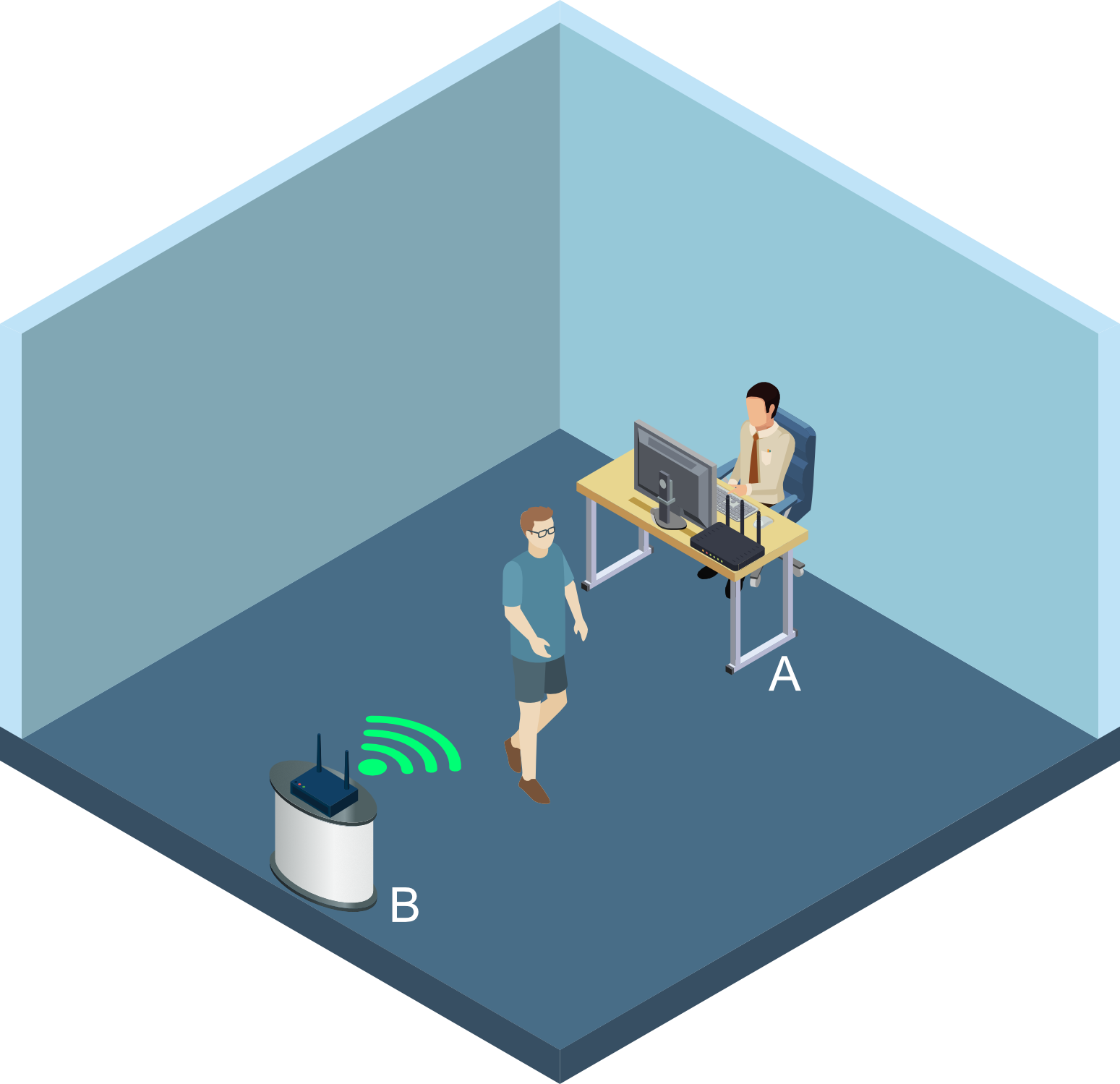}
	\caption{Proposed method architecture. (A) Desktop PC, (B) 802.11n enabled commercial router.}\label{fig:architecture}
\end{figure}

\section{Proposed Method}\label{03}
In this section the architecture of the proposed method is reported in detail beginning with hardware components followed by underlying technologies, CSI measurement, SNR estimation, and learning and re-identification strategies.

\subsection{Method Architecture}
Our proposed method architecture (Fig. \ref{fig:architecture}) is designed to be simple, in fact, it is comprising of one desktop PC (Fig. \ref{fig:architecture}A) and one 802.11n commercial router (Fig. \ref{fig:architecture}B). The PC, which is the receiver, has one Intel Wi-Fi Link 5300 (IWL5300) Network Interface Card (NIC) with three antennas mounted on it. The router, which is the transmitter, has two antennas and the required 802.11n protocol enabled for transmissions. The architecture exploits the Multiple-Input and Multiple-Output (MIMO) technology, thus both transmitter and receiver make use of all of their available antennas for transmission and reception tasks. When a person is either between or passing through the transmitting and receiving locations, Wi-Fi signals reflection is affected by its presence with respect to the case in which there is an unobstructed path. The CSI measured on received signals has characteristics that could be used to extract features for training a model able to re-identify people. In the proposed method, given a set $U$ of person identities, we exploit CSI samples measured from OFDM subcarriers to extract SNR (i.e., the ratio of signal power to noise power) as feature for people Re-ID (Fig. \ref{fig:snr}) by using a Multi-Layer Perceptron (MLP) neural network. 

\subsection{Orthogonal Frequency-Division Multiplexing}
The orthogonal frequency-division multiplexing (OFDM) is a method used in modern wireless communications for digital data encoding on multiple carrier frequencies. It provides improvement in terms of communication performance by exploiting frequency diversity of the communication channels. In recent years, this technology is used in popular wireless networks, including IEEE $802.11 a/g/n$ and $4G$. Data is divided into multiple streams, each of them coded and modulated on different subcarriers on adjacent frequencies. Generally, overlapping adjacent channels can interfere with one another. To avoid this, in OFDM each subcarrier is orthogonal to each other in order to minimize interference during transmissions, thus the maximum power of each sub-carrier corresponds directly with the minimum power of each adjacent channel. For example, for the OFDM used by $802.11n$ physical layer, a 20 or 40 MHz channel is composed of 56 or 114 subcarriers, respectively, such that each subcarrier can be used as a narrowband channel. This is why we chose to use the Channel State Information measured from OFDM subcarriers, it provides a finer granularity of the channel state useful in achieving higher accuracy for Re-ID in practice. 

\subsection{Channel State Information}
In MIMO-based technology systems, multiple transmitting and receiving antennas are used to take advantage from multi-path propagation. Formally, the system can be modeled as:

\begin{equation}
	y = Hx + n
\end{equation}

where $y$ is the received signal $n \times 1$ vector, $x$ is the transmitted signal $n \times 1$ vector, $H$ and $n$ are the channel matrix and noise $n \times 1$ vector, respectively. The channel matrix $H$, with $m$ antennas for transmission and $n$ antennas for reception, can be defined as:

\begin{equation}
	H = \begin{vmatrix}
		h_{1,1} & h_{1,2} & \hdots & h_{1,m}\\
		\vdots & \vdots & \vdots & \vdots\\
		h_{n,1} & h_{n,2} & \hdots & h_{n,m}\\
	\end{vmatrix}
\end{equation}

where $h_{n,m}$ is the gain of each path between the $m-th$ transmitter and the $n-th$ receiver. Basically, there are two types of CSI, i.e., instantaneous and statistical ones. The former is estimated in fast fading systems where channel conditions vary rapidly during transmission. On the contrary, the latter can be only estimated in slow fading systems.
Theoretically, if H is known, considering the channel estimation errors the instantaneous CSI can be modeled as:

\begin{equation}
	vec(H_{estimate}) \sim \textit{CN}(vec(H),R_{error})
\end{equation}

where $H_{estimate}$ is the channel estimation, $R_{error}$ is the covariance matrix of the estimation error. Since the conditions of channel H vary, the instantaneous CSI is estimated on short-time basis. Channel matrix H is estimated by combining knowledge of both transmitted and received signals. Given a sequence $p_1, \hdots, p_N$, where each vector $p_i$ is transmitted over the channel as:

\begin{equation}
	y_i = Hp_i + n_i
\end{equation}

by combining $y_i$ (i.e., received signals) with transmitted signals $P = [p_1, \hdots, p_N]$ and noise $N = [n_1, \hdots, n_N]$ matrices, the total signalling becomes:

\begin{equation}
	Y =[y_1,\ldots ,y_N] = HP + N
\end{equation}

and the CSI can be recovered from the knowledge of $Y$ and $P$.

\subsection{Signal Noise Ratio Estimation}
\begin{figure}[!t]
	\centering
	\subfigure[]{\includegraphics[width=0.4\textwidth]{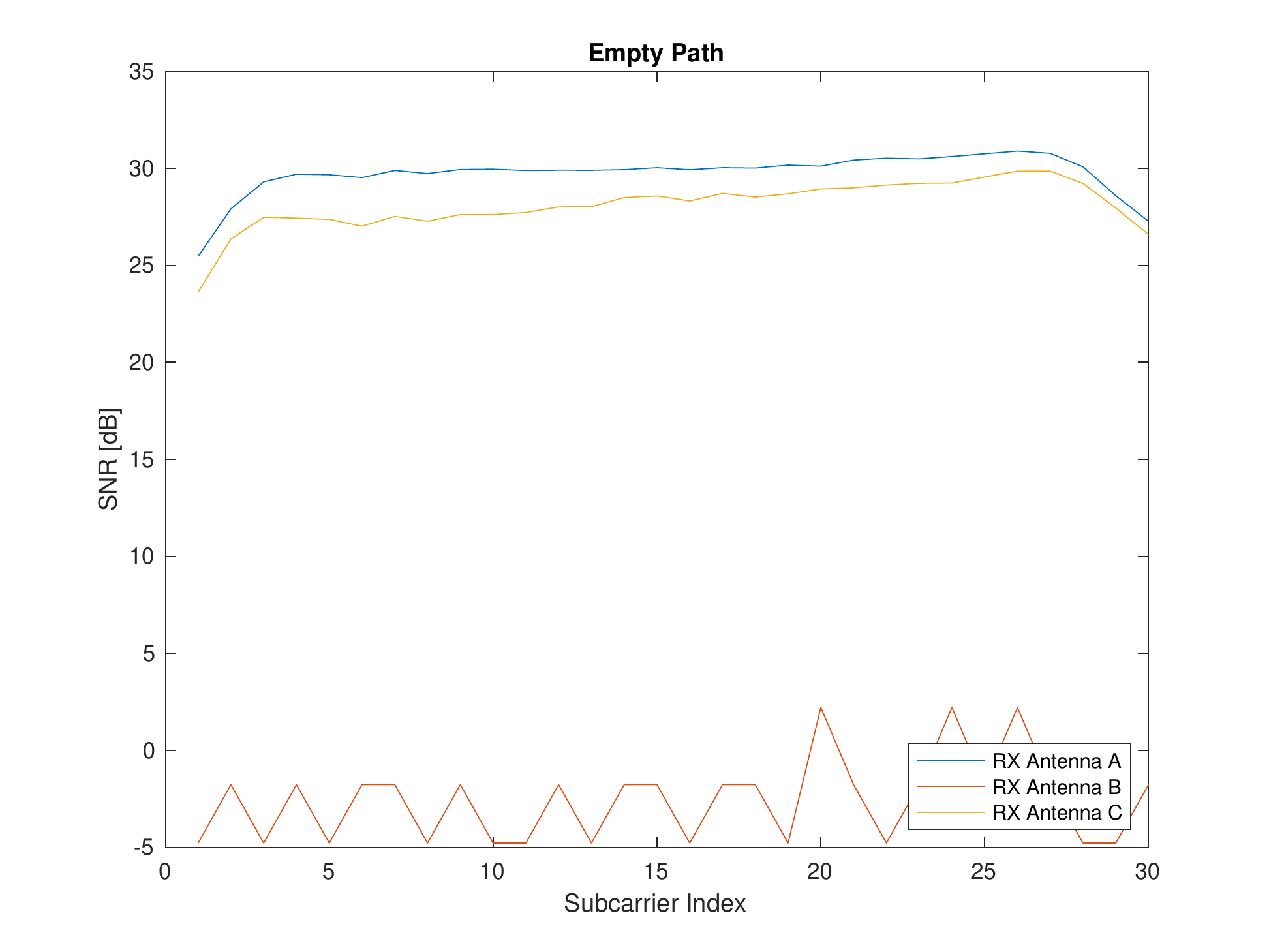}}
	\subfigure[]{\includegraphics[width=0.4\textwidth]{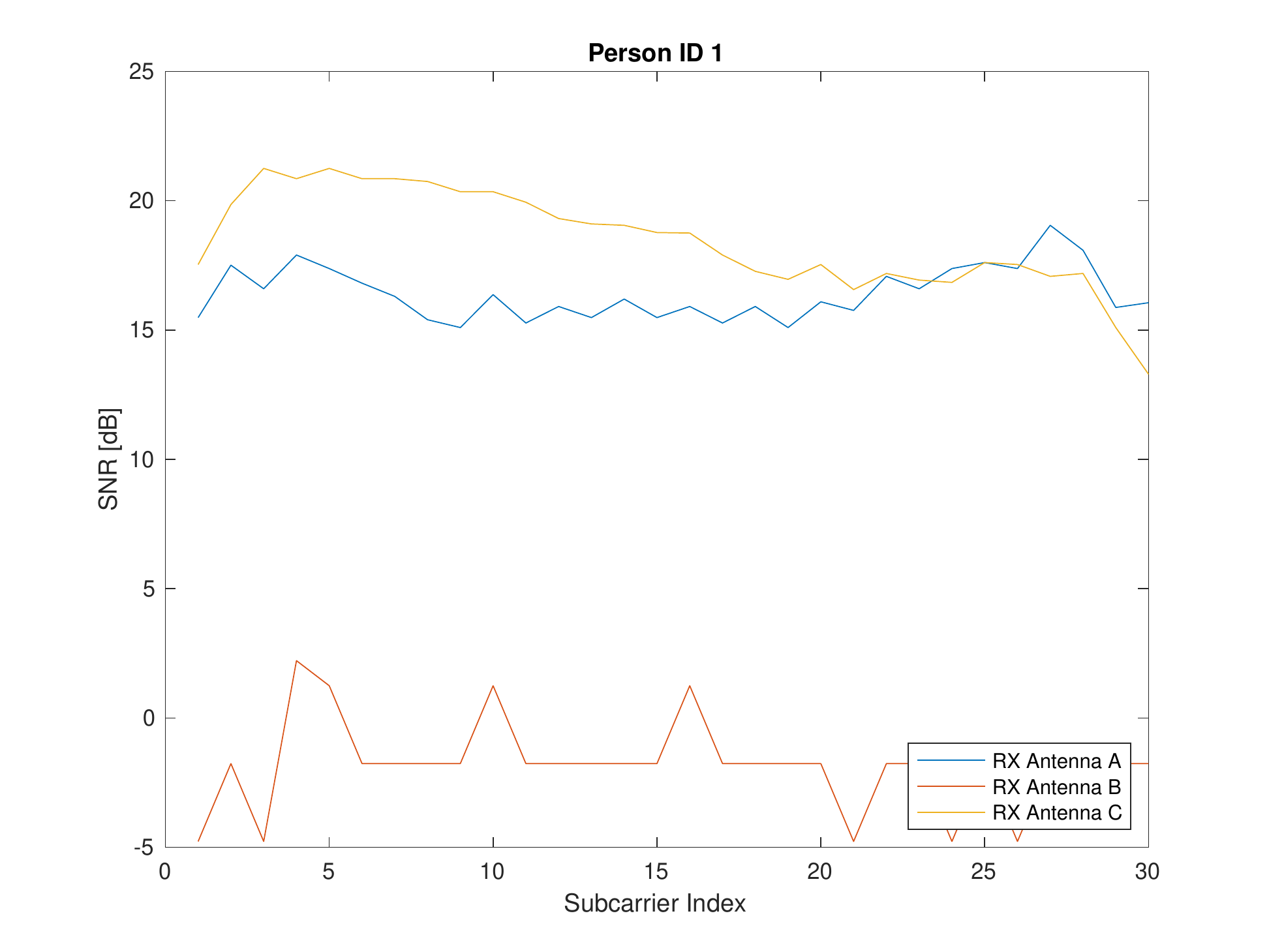}}
	\subfigure[]{\includegraphics[width=0.4\textwidth]{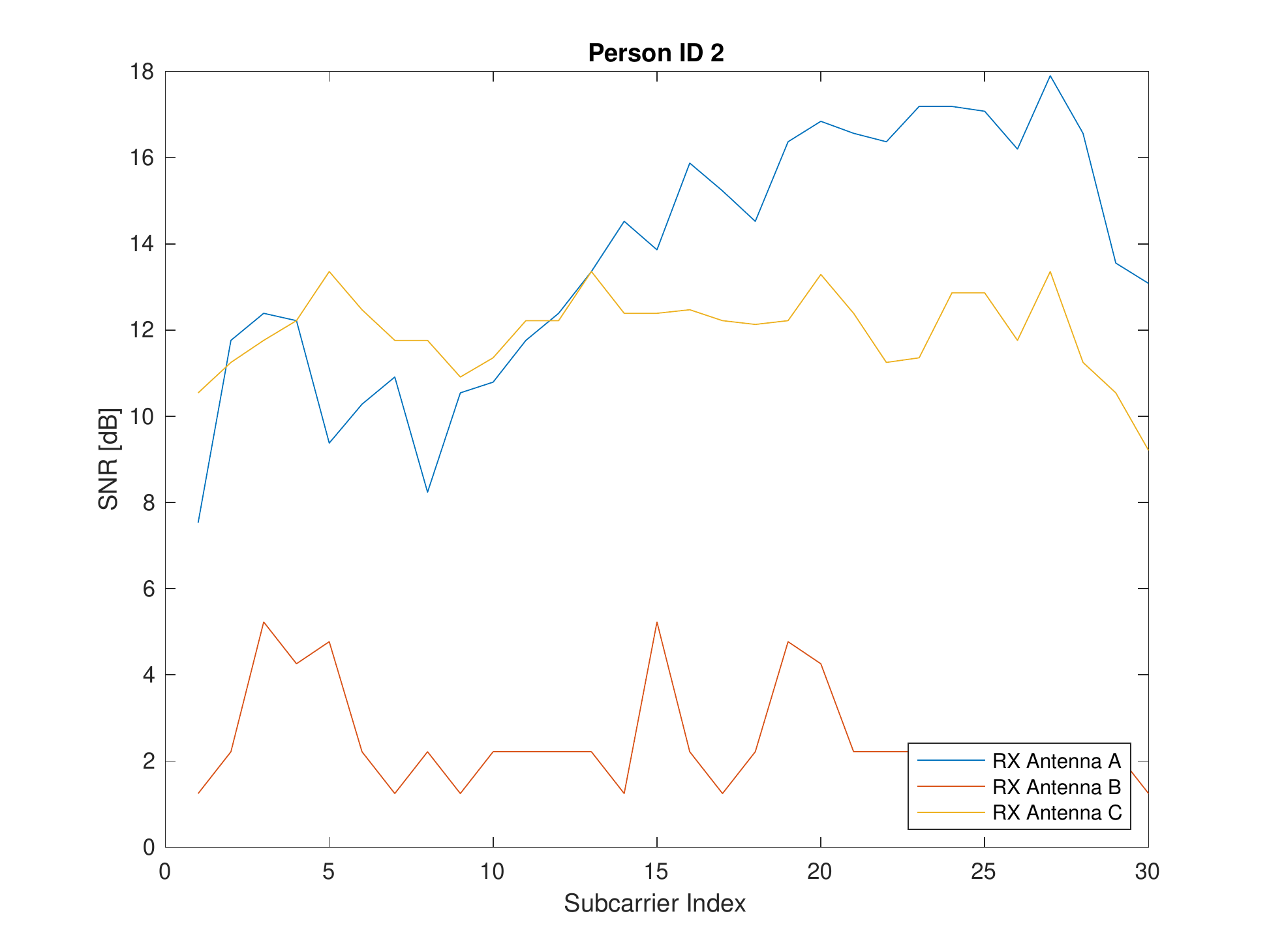}}
	\subfigure[]{\includegraphics[width=0.4\textwidth]{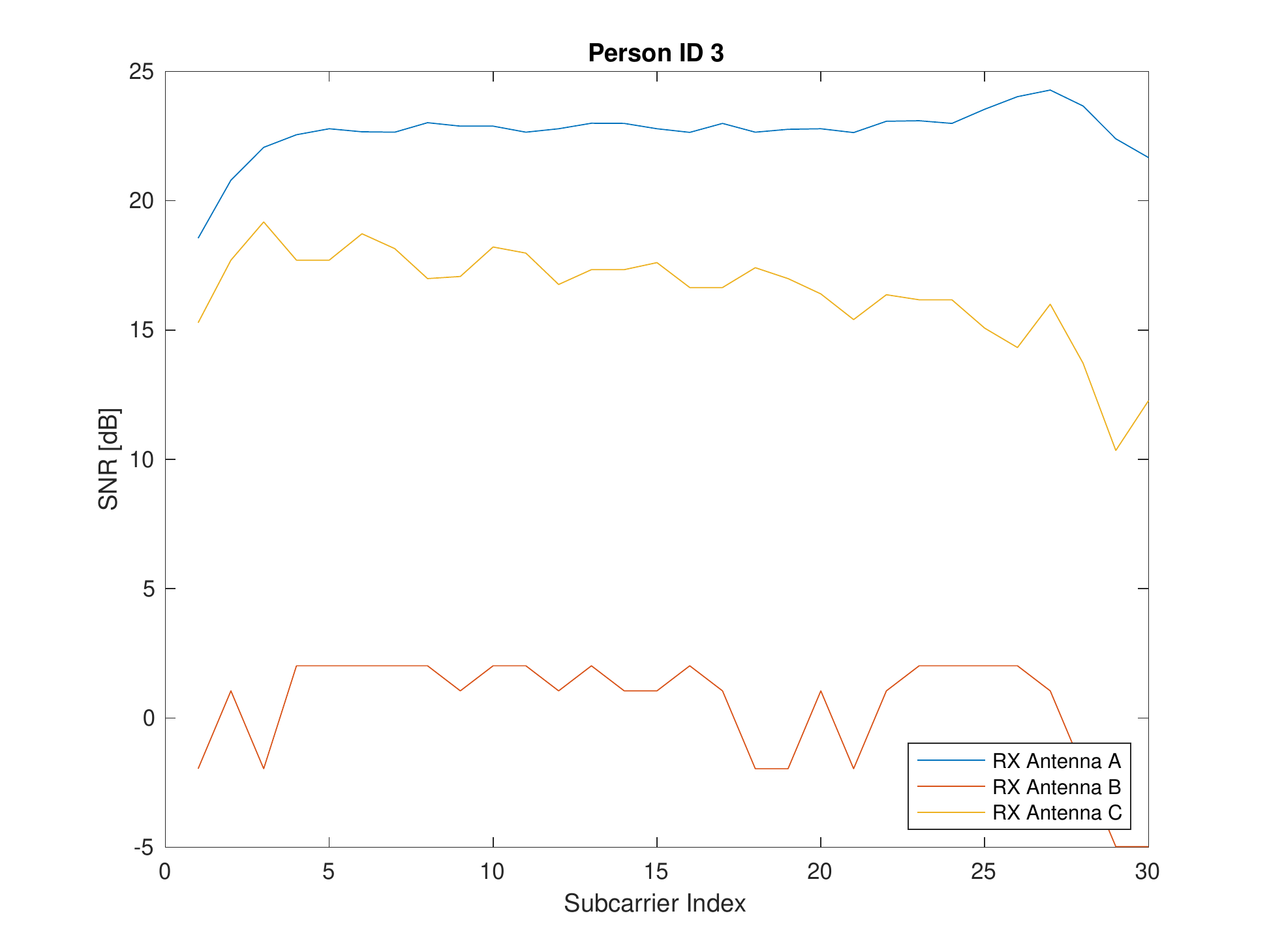}}
	\caption{SNR examples with 3 receiving antennas.}
	\label{fig:snr}
\end{figure}

In wireless transmissions, radio signals propagation paths and characteristics change accordingly the environment and objects encountered before arriving to the receiver. For CSI measurement, the network interface card used in the proposed method is the aforementioned IWL5300. Since it was designed for commercial use, custom firmware and driver versions of the ones used by \cite{Halperin11} were required. Specifically, we measure the channel state in communications between the Access Point (i.e., the router) and the NIC. The measured CSI, consisting of complex-valued and high-dimensional channel matrices for 30 subcarriers, makes difficult achieving the person Re-ID task. Therefore, given a person identity $u \in U$ , the channel state of N packets from $u$ is measured. Then, obtained the CSI samples, we extract the Signal Noise Ratio per packet as a k-dimensional vector $SNR_u$ defined as:

\begin{equation}
	SNR_u^{i,j} = 1_{1\times K} \frac{1}{K} \sum_{k=1}^K SNR_u(k)
\end{equation}

where $K$ is the number of subcarriers within each CSI sample, and $1 \times K$ is the K-lenght all-one SNR vector. Because the wireless devices are equipped with multiple antennas, the SNR is computed from each communication channel state between the transmitting antenna $i$ and the receiving antenna $j$. In this way, the high-dimension complex-valued person biometrics within the measured CSI are mapped into the SNR, therefore the feature dimension is reduced. For example, Fig. \ref{fig:snr} shows the SNR, from different packets, related to empty path between transmitting and receiving locations (Fig. \ref{fig:snr} a) or person identity (Fig. \ref{fig:snr} b,c,d).

\subsection{Multi-Layer Perceptron}
Both learning and re-identification steps are based on the use of a Multi-Layer Perceptron (MLP) \cite{Bishop:1995:NNP:525960} network consisting of two hidden layers with LeakyReLU \cite{maas2013rectifier} as activation function. The latter has been chosen because the experiments shown that it does not suffer from "dying ReLU" \cite{lu2019dying} problem and speeds up the training. The network loss function is the Cross-Entropy one, and the activation function for the last layer is the Soft-max one. The optimization algorithm used for the network is Adam \cite{adam}, because empirical results demonstrate achieving of good results faster. Moreover, activations in hidden layers are normalized by using Batch Normalization \cite{ioffe2015batch} to improve accuracy. For each person identity $u \in U$, the input of the network are the SNR vector estimated from $N$ packets related to $u$. During the learning stage, for each identity more packets are used to learn its fingerprint, whereas during testing stage even only one packet can be used for the Re-ID. The latter shows that our method is suitable for real-time applications.

\section{Experiments}\label{04}
In this section the performed experiments are reported. In detail, the datasets and the obtained results are carefully discussed.

\subsection{Datasets}
To the best of our knowledge there are no datasets containing Wi-Fi signals concerning person identities, therefore we acquired our own datasets to evaluate the proposed method performance. To collect the data, the following acquisition protocol has been used. A total of 50 person identities have been acquired in 4 different conditions: standing still facing the router, standing still not facing the router, passing through the PC and the router from left to right and, finally, passing through the PC and the router from right to left. These acquisitions have been organized in two datasets, i.e., the standing and the walking ones. Three different acquisitions per subject have been collected for each of aforementioned conditions. This allows the model to capture all changes in both standing and walking patterns for each subject. Moreover, each acquisition lasts 3 seconds with a maximum number of 200 packets transmitted and received. Both the duration of acquisitions and the number of packets to use have been found empirically during preliminary tests. In addition, the acquisition of empty path between transmitting and receiving locations has been performed in order to reduce false positives. Finally, data has been collected in normal rooms, i.e., without using any shielding mechanism for interference introduced by other devices such as smartphones and other Wi-Fi devices. This kind of noisy acquisitions allowed us to test the robustness of the proposed approach for real applications.

\subsection{Results}
\begin{figure}[t]
	\centering
	\subfigure[]{\includegraphics[width=0.45\textwidth]{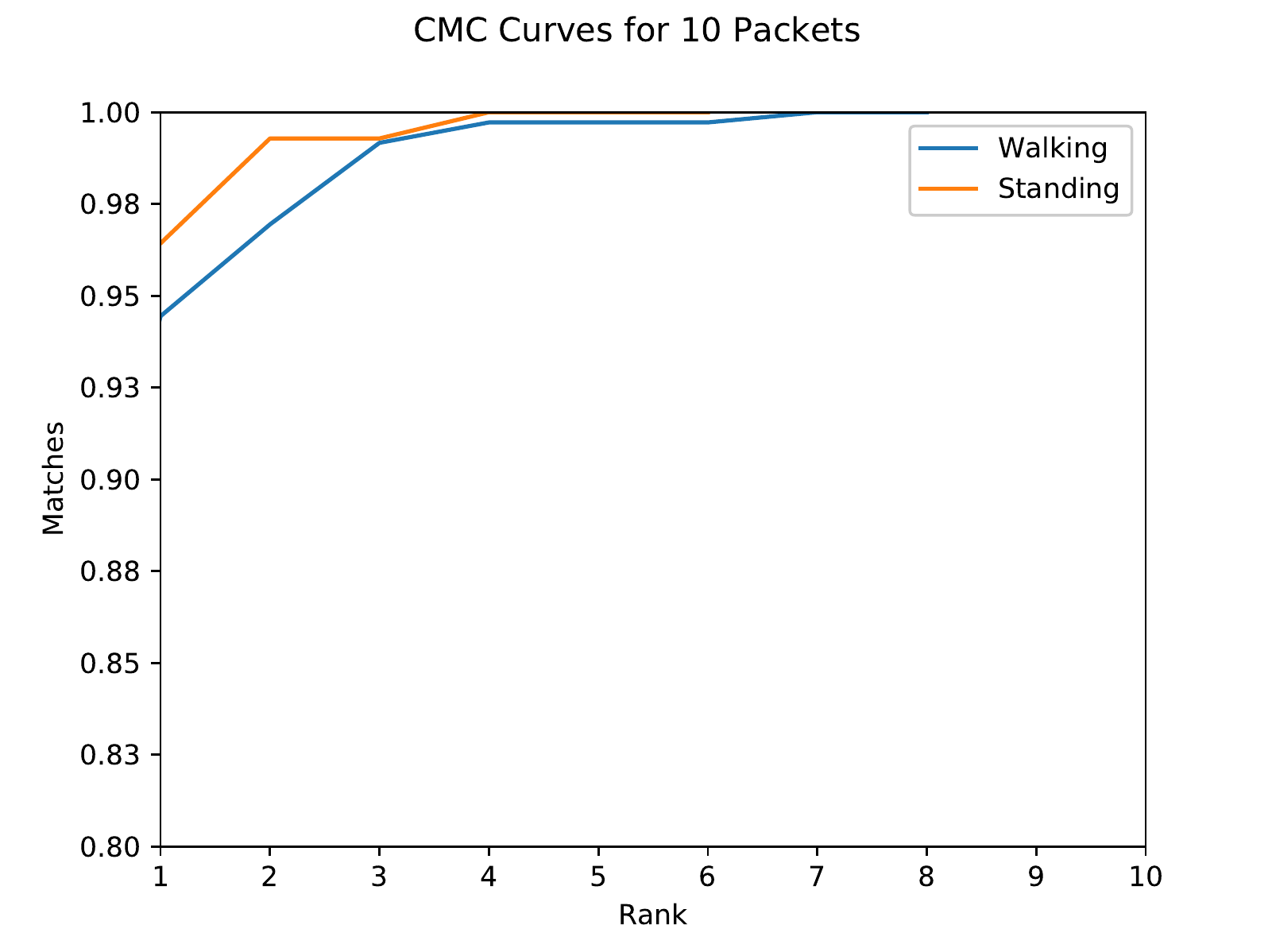}}
	\subfigure[]{\includegraphics[width=0.45\textwidth]{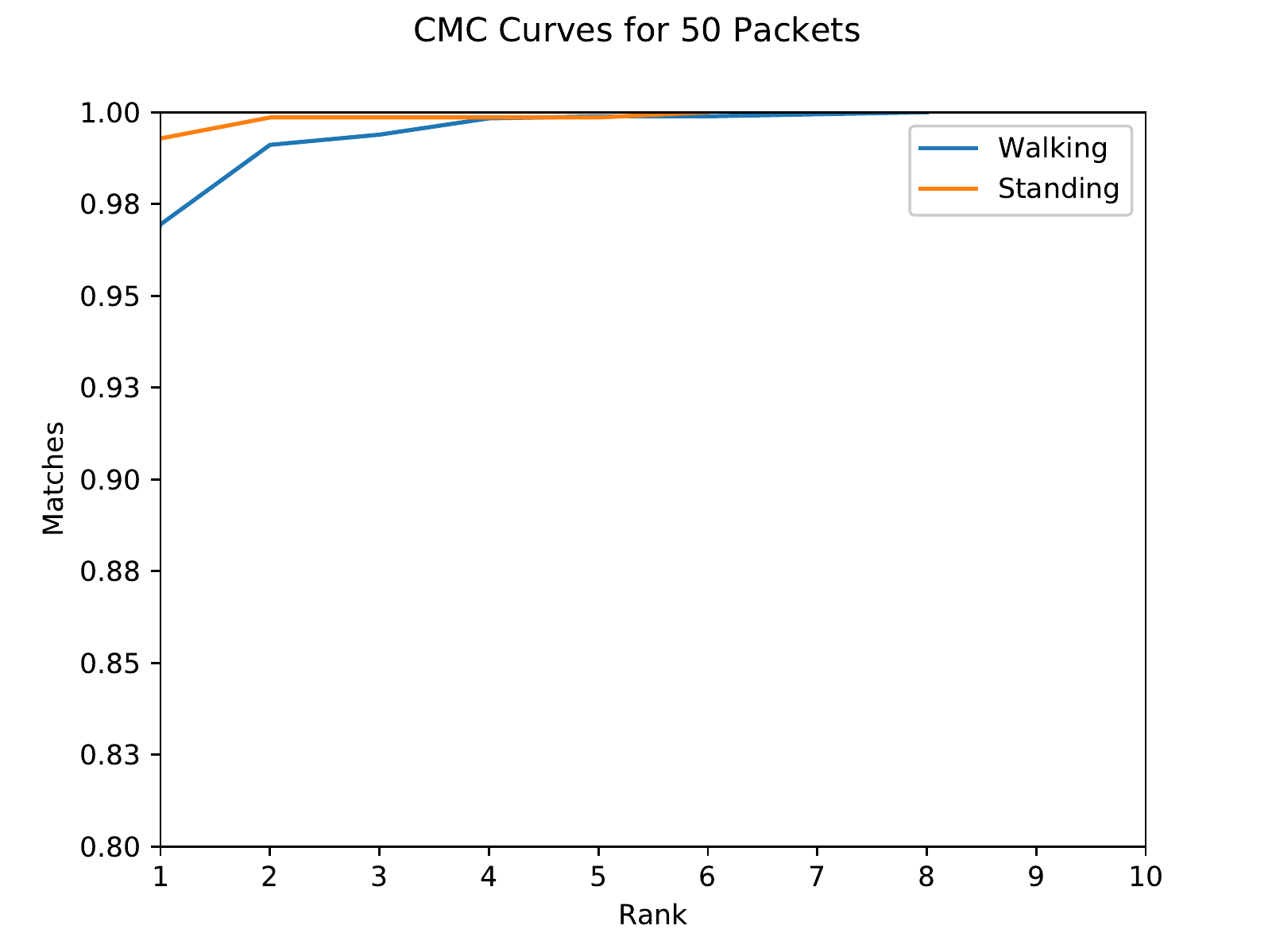}}
	\subfigure[]{\includegraphics[width=0.45\textwidth]{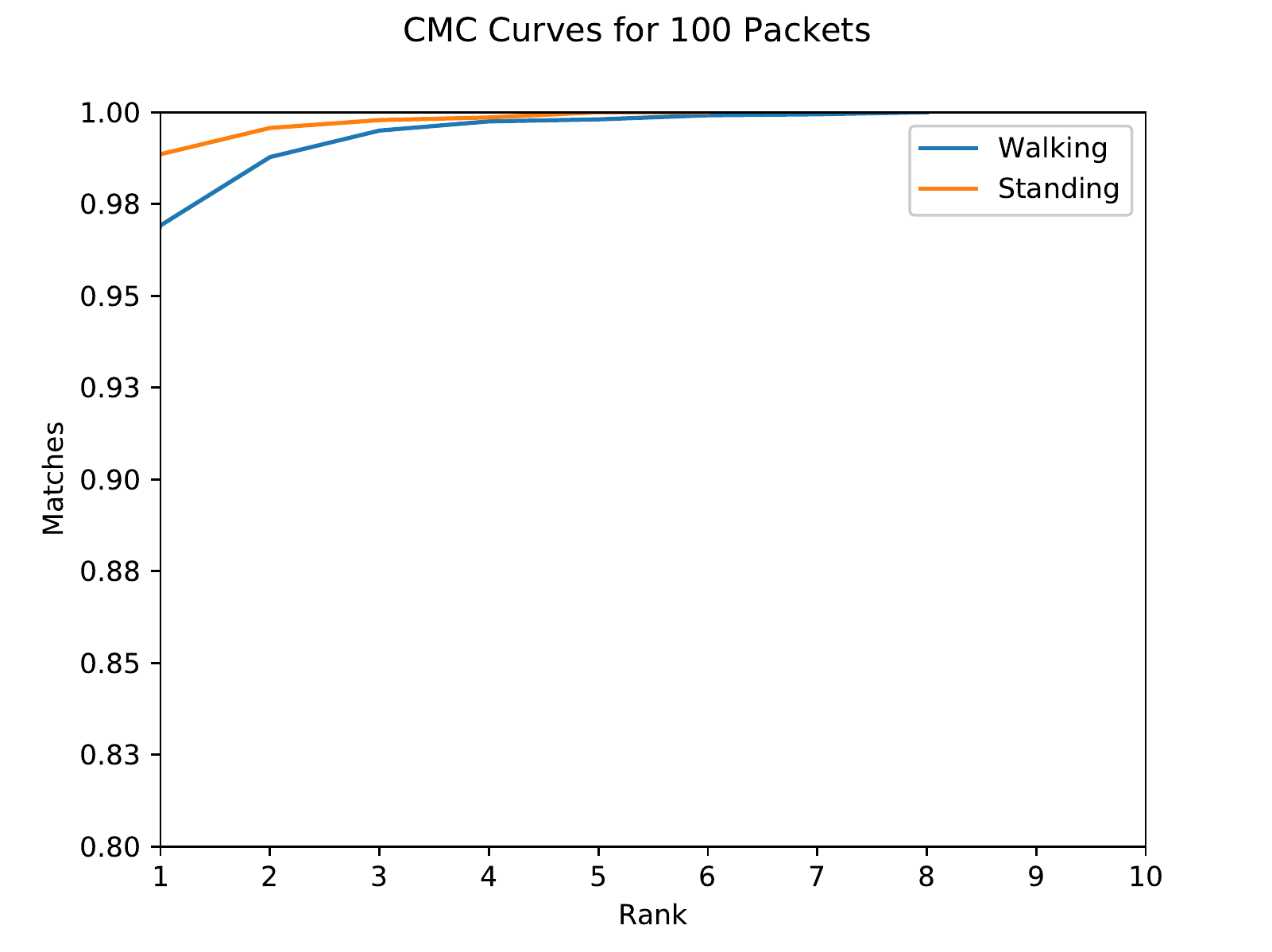}}
	\subfigure[]{\includegraphics[width=0.45\textwidth]{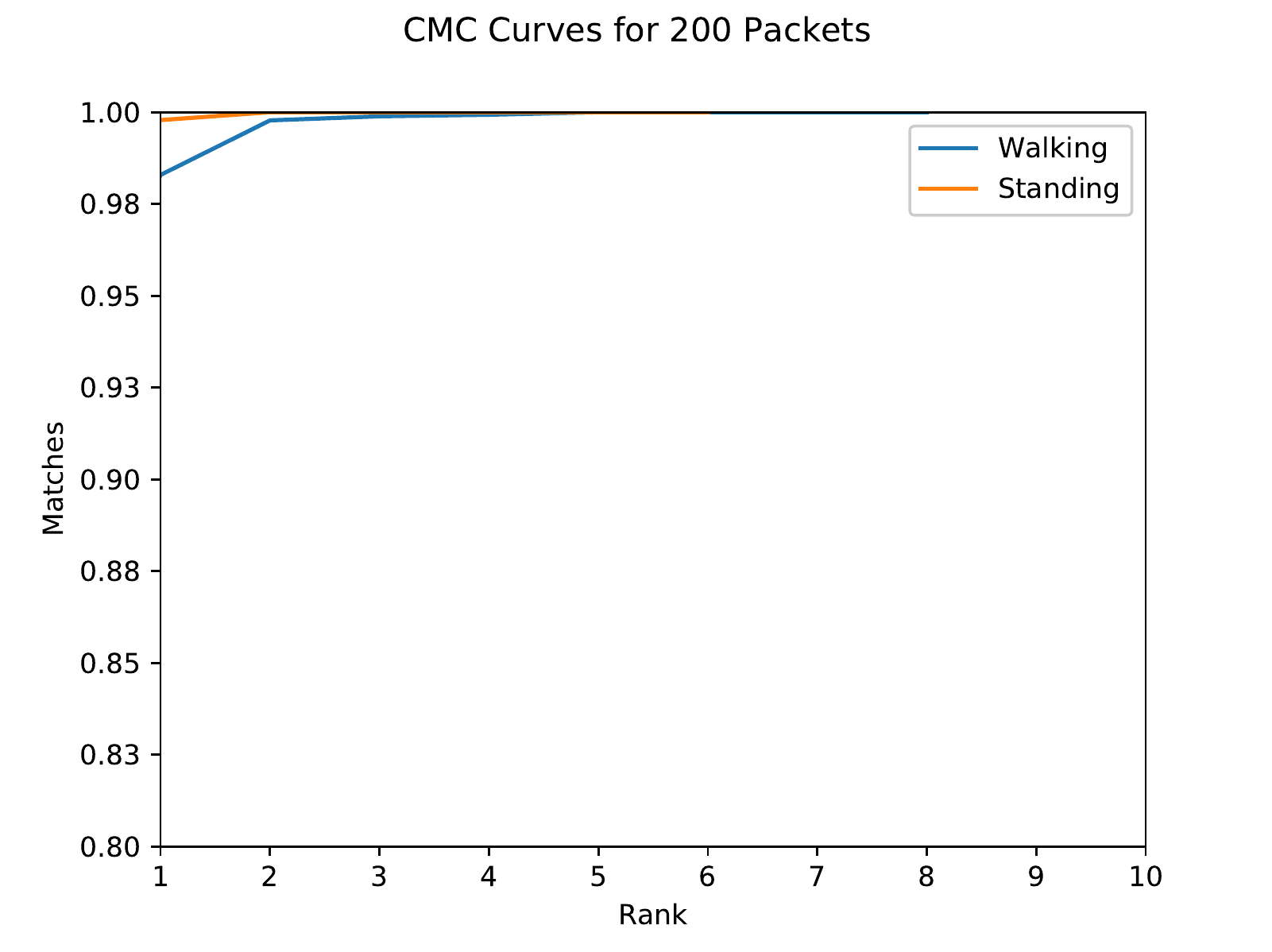}}
	\caption{CMC curves for re-identification by using a) 10 packets, b) 50 packets, c) 100 packets, and d) 200 packets.}
	\label{fig:cmc}
\end{figure}

In Table \ref{tab:packets_result}, the results obtained by using both the two datasets, considering different packets number per identity, are shown. We used a minimum of 10 and a maximum of 200 packets. This because using less of 10 packets led to bad performance, while using a number of packets higher than 200 does not really increases the model accuracy. Concerning subcarriers, we have used all the 30 subcarriers provided by the network interface.
Obviously, better results have been obtain by using the maximum number of packets available, leading to obtain a validation accuracy of $97.86\%$ for the standing dataset and $91.11\%$ for the walking one. By using only 10 packets, instead, we obtained a validation accuracy of $93.57\%$ for the standing dataset and $82.22\%$ for the walking one.
\begin{table}
	\centering
	\caption{Table showing accuracies obtained with respect to the number of packets and the dataset used.}
	\begin{tabular}{c c c c c c}
		\hline
		\hline
		\textbf{No. of Packets} & \textbf{No. of subcarriers} & \textbf{Dataset} & \textbf{Training Accuracy} & \textbf{Validation Accuracy} & \textbf{Rank 1} \\ \hline
		10 & 30 & Walking & 98.30\% & 82.22\% & 82\%\\
		10 & 30 & Standing & 98.13\% & 93.57\% & 93\% \\\hline \hline
		50 & 30 & Walking & 93.42\% & 88.28\% & 88\% \\
		50 & 30 & Standing & 96.56\% & 96.43\% & 96\% \\\hline \hline
		100 & 30 & Walking & 94.71\% & 88.80\% & 88\% \\
		100 & 30 & Standing & 96.98\% & 96.5\% & 96\% \\\hline \hline
		200 & 30 & Walking & 96.79\% & 91.11\% & 91\%\\
		200 & 30 & Standing & 98.07\% & 97.86\% & 97\%\\
		\hline
		\hline
	\end{tabular}
	\label{tab:packets_result}
\end{table}
In general, the experiments performed on the standing dataset gives better results. This may be due to the fact that the moving subjects in the walking dataset could suffer of the Doppler shift. This aspect will be further investigated in order to increase the performance of our method. By looking at our Rank 1 best value, i.e. $97\%$, we outperform most of the works at the current state-of-the-art which use visual features, remarking the goodness of the proposed approach. In Figure \ref{fig:cmc}, the Cumulative Match Curve (CMC) for the different packets number are depicted. As it is possible to notice, our method achieves rapidly a $100\%$ match score in the first 10 ranks, which is quite impressive considering the proposed approach.

\section{Conclusion}\label{05}
In this paper, a Wi-Fi passive re-identification system is proposed. The proposed method uses a commercial Wi-Fi network interface card to extract CSI from the received signals. A Multilayer Perceptron network trained with SNR estimated from CSI measurements is used to perform the person re-identification task. To the best of our knowledge there are no publicly available datasets concerning the re-identification through Wi-Fi signals, therefore we created two datasets comprising of 50 standing and walking human identities between transmitting and receiving locations. In our experiments, we have obtained a rank 1 accuracy of $97\%$ which is a very promising result.

\bibliographystyle{unsrt}  
\pagebreak

\end{document}